\documentclass[preprint,showpacs,preprintnumbers,amsmath,amssymb,showkeys]{revtex4}
\usepackage{graphicx}% Include figure files
\usepackage{dcolumn}% Align table columns on decimal point
\usepackage{bm}% bold math

\begin{document}

\preprint{}

\title{Role of transverse momentum currents in optical Magnus effect in the free space}% Force line breaks with \\
\author{Hailu Luo}
%\altaffiliation[ ]{}%Lines break automatically or can be forced with \\
\author{Shuangchun Wen}\email{scwen@hnu.cn}
\author{Weixing Shu}
\author{Dianyuan Fan}
%%\email{Second.Author@institution.edu}
\affiliation{Key Laboratory for Micro/Nano Opto-Electronic Devices
of Ministry of Education, Hunan University, Changsha 410082, China}
\date{\today}% It is always \today, today,

\begin{abstract}
We establish a general vector field model to describe the role of
transverse momentum currents in optical Magnus effect in the free
space. As an analogy of mechanical Magnus effect, the circularly
polarized wavepacket in our model acts as the rotating ball and its
rotation direction depends on the polarization state. Based on the
model we demonstrate the existence of a novel optical
polarization-dependent Magnus effect which is significantly
different from the conventional optical Magnus effect in that
light-matter interaction is not required. Further, we reveal the
relation between transverse momentum currents and optical Magnus
effect, and find that such a polarization-dependent rotation is
unavoidable when the wavepacket possesses transverse momentum
currents. The physics underlying this intriguing effect is the
combined contributions of transverse spin and orbital currents. We
predict that this novel effect may be observed experimentally even
in the propagation direction. These findings provide further
evidence for the optical Magnus effect in the free space.
\end{abstract}


%Use showkeys class option if keyword
\maketitle

\section{Introduction}\label{SecI}
The optical Magnus effect is the photonic version of the Magnus
effect in classical mechanical systems. As a result of the usual
mechanical Magnus effect, a rotating ball falling through the air is
deflected from the vertical to the direction of rotation. The
influence of spin on the trajectory can be considered as the optical
analogue of the mechanical Magnus effect. The spin photon to some
extent can be regarded as a rotating ball, and then after multiple
reflections during its propagation through a fiber the photon will
be deflected from its initial
trajectory~\cite{Dooghin1992,Liberman1992}. Hence this effect has
received the name of optical Magnus effect or optical ping-pong
effect. Recently, the physical nature of the optical Magnus effect
is connected with spin-orbital interaction in a wave field, which
essentially depends on the gradient of a refractive index of optical
medium~\cite{Bliokh2008}.

For beams propagating in the free space, it is generally believed
that the optical magnus effect takes no place since the gradient of
a refractive index is zero. In fact, the changes in transverse
momentum currents can lead to the rotation phenomenon in paraxial
light beams. The possible occurrence of a polarization-dependent
rotation of beam centroid in the free space was already predicted by
Borodavka and coworkers~\cite{Borodavka1999}. However, they do not
connect such a rotation with the occurrence of nonzero transverse
momentum currents. In our opinion, the optical Magnus effect
manifests itself within vector field structure in the frame of
classical electrodynamics. However, the Jones vector is not
sufficient to describe the vectorial property of a finite beam, due
to the longitudinal component~\cite{Pattanayak1980}. Thus, it is
necessary for us to establish a general vector field model to reveal
the role of transverse momentum currents in optical Magnus effect.

Meanwhile the same interaction also leads to other effects such as
the spin Hall effect of light (SHEL)~\cite{Onoda2004,Bliokh2006}.
The interesting effect has been recently observed in beam
refraction~\cite{Hosten2008} and in scattering from dielectric
spheres~\cite{Haefner2009}. More recently, the SHEL was found to
occur when a light beam is observed on the direction making an angle
with the propagation axis~\cite{Aiello2009a}. This effect has a
purely geometric nature and amounts to a polarization-dependent
shift or split of the beam intensity distribution. As the tilting
angle tends to zero, the splitting effect vanishes. The possible
reason for this is that the transverse intensity distribution of
fundamental Gaussian beam is axially symmetric and the
polarization-dependent rotation, if it exists, cannot be observed
directly. We believe that such a polarization-dependent shift should
be unavoidable when the beam possesses the transverse momentum
currents. Thus, our another motivation is to prove the intriguing
optical Magnus effect can be observed even in the beam propagation
direction.

In this work, we want to explore what role of the transverse
momentum currents play in the optical Magnus effect in the free
space. First, we introduce the Whittaker scale potentials to
describe the vector field structure for different polarization
models. In the frame of classical electrodynamics, it is the
rotating wavepacket but not the spin photon acting as the spin ball,
which is significantly different from the previous works. Then, we
uncover how the centroid of wavepacket evolves, and how the
polarization state affects its rotation of the centroid trajectory.
Finally, we examine what roles the spin and orbital currents play in
the optical Magnus effect. A relation between transverse momentum
currents and optical Magnus effect is obtained. We believe that our
findings may provide insights into the fundamental properties of
optical currents in beam propagation.

\section{Vector field model}\label{SecII}
\begin{figure}
\includegraphics[width=12cm]{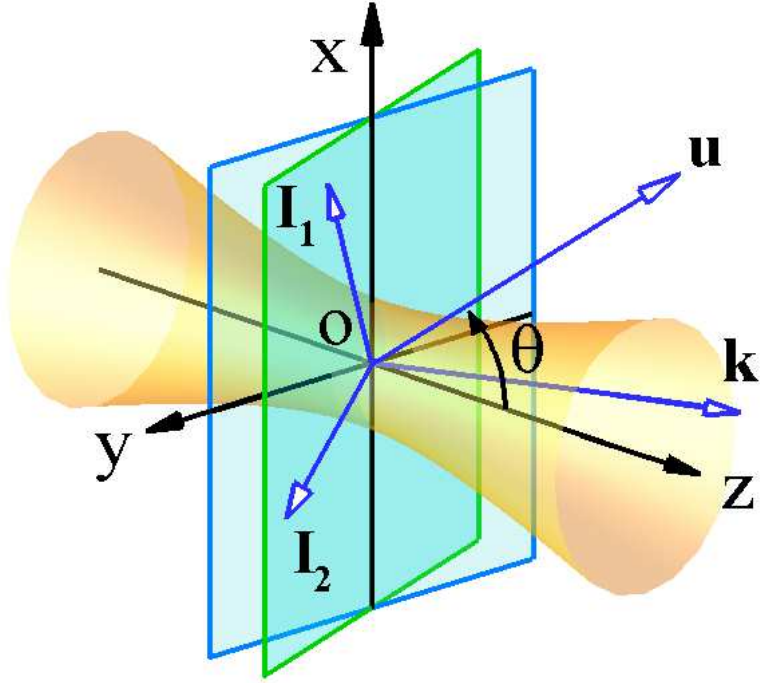}
% Here is how to import EPS art
\caption{\label{Fig1} (color online) Schematic illustrating the
geometry of the vector field structure. The fixed unit vector
$\mathbf{u}$ lies in the plane $zox$ and makes an angle $\theta$
with the propagation axis $z$. The propagation axis $z$
perpendicular to the plane of $xoy$, and the wavevector $\mathbf{k}$
for an arbitrary angular spectrum component is perpendicular to the
plane of $I_1oI_2$ }
\end{figure}

In order to reveal the role of the transverse momentum currents in
optical Magnus effect, we begin to establish a general beam
propagation model to describe the vector field structure.
Figure~\ref{Fig1} illustrates the geometry of the vector field
structure in the Cartesian coordinate system. Following the standard
procedure, the electric field $\mathbf{E}(\mathbf{r},\omega)$ is
obtained by solving the vector Helmholtz equation
\begin{equation}
(\nabla^2+k^2)\mathbf{E}(\mathbf{r},\omega)=0,\label{VHE}
\end{equation}
where $\mathbf{r}=x \mathbf{e}_{x} +  y \mathbf{e}_{y}+ z
\mathbf{e}_{z}$ and $k=\omega/c$ is the wave number in the free
space. The vector Helmholtz equation can be solved by employing the
angular spectrum representation of the electric field as
\begin{eqnarray}
\mathbf{E}(\mathbf{r})&=&\int d k_{x}dk_{y}
\tilde{\mathbf{E}}(k_{x},k_{y})\nonumber\\
&&\times\exp [i(k_{x}x+k_{y}y+ k_{z} z)]\label{ASR}.
\end{eqnarray}
The transverse condition
$\nabla\cdot\tilde{\mathbf{E}}(k_{x},k_{y})= 0$ implies that the
components of the angular spectrum satisfy the relation
$\mathbf{k}\cdot\tilde{\mathbf{E}}(k_{x},k_{y})=0$. In principle,
due to the longitudinal component, the Jones vector is not
sufficient to describe the vectorial properties of a finite beam.
Hence it is necessary for us to define two mutually orthogonal
vectors
\begin{equation}
\mathbf{I}_1=\frac{\mathbf{I}_2\times\mathbf{k}}{|\mathbf{I}_2\times\mathbf{k}|},~~~~~
\mathbf{I}_2=\frac{\mathbf{k}\times\mathbf{u}}{|\mathbf{k}\times\mathbf{u}|},
\end{equation}
describing the vector field structure. Here, the fixed unit vector
$\mathbf{u}$ lies in the plane $zox$ and makes an angle $\theta$
with the propagation axis, and
\begin{equation}
\mathbf{u}=\sin\theta\mathbf{e}_x+\cos\theta\mathbf{e}_z\label{FUV},
\end{equation}
where $-\pi/2 \leq\theta\leq +\pi/2$. Note that this fixed unit
vector is not a purely mathematical concept~\cite{Li2009a,Li2009b}.
In fact, a perpendicular $\mathbf{u}$ to the propagation axis
corresponds to the uniformly polarized beam in the paraxial
approximation~\cite{Davis1979}, and a parallel $\mathbf{u}$
corresponds to the cylindrical vector beam~\cite{Davis1981}. The
axis $\mathbf{u}$ that is neither perpendicular nor parallel to the
propagation axis was observed by Hosten and Kwiat in a recent
experiment~\cite{Hosten2008}.

In order to accurately describe the optical Magnus effect, we
introduce the scaler Whittaker potentials~\cite{Pattanayak1980} to
represent the vectorial field. Consequently, the angular spectrum
can be decomposed along the two vectors and we have
\begin{equation}
\tilde{\mathbf{E}}(k_{x},k_{y})=\tilde{V}_1(k_{x},k_{y})
\mathbf{I}_1+\tilde{V}_2(k_{x},k_{y})\mathbf{I}_2\label{vector}.
\end{equation}
Here $\tilde{V}_j=(\alpha~~~\beta)^T \tilde{f}(k_{x},k_{y})$
($j=1,2$) are the scaler Whittaker potentials. The amplitude of the
angular spectrum is referred to as
\begin{equation}
\tilde{f}(k_{x},k_{y})=\frac{w_0}{\sqrt{2\pi}}\exp\left[-\frac{k_x^2+k_y^2}{4}w_0^2\right],
\end{equation}
where $w_0$ is beam-waist size. The matrix $(\alpha~~~\beta)^T$
denotes the Jones vector which satisfies the normalization condition
$\alpha\beta^\ast+\alpha^\ast\beta=1$. The coefficients $\alpha$ and
$\beta$ satisfy the relation
$\sigma=i(\alpha\beta^\ast-\alpha^\ast\beta)$. The polarization
operator $\sigma=\pm1$ corresponds to left and right circularly
polarized light, respectively. It is well known that a circularly
polarized beam can carry spin angular momentum $\sigma\hbar$ per
photon due to its polarization state~\cite{Beth1936}.

From the viewpoint of Fourier optics~\cite{Goodman1996}, the
Whittaker potentials $V_j$ are given by the relation
\begin{eqnarray}
V_j(\mathbf{r})&=&\frac{1}{k^2}\int d k_{x}dk_{y}
\tilde{V}_j(k_{x},k_{y})\nonumber\\
&&\times\exp [i(k_{x}x+k_{y}y+ k_{z} z)]\label{nopr}.
\end{eqnarray}
It can be verified that both $V_1$ and $V_2$ satisfy the scalar
Helmholtz equation:
\begin{equation}
(\nabla^2+k^2)V_j(\mathbf{r})=0\label{SHE}.
\end{equation}
On substituting Eq.~(\ref{vector}) into Eq.~(\ref{ASR}), the
electric field can be expressed in terms of the Whittaker potentials
\begin{eqnarray}
\mathbf{E}(\mathbf{r})&=&\nabla\times\nabla\times
(\mathbf{u}V_1)-ik\nabla\times(\mathbf{u}V_2)\label{FieldE}.
\end{eqnarray}
In fact, after the angular spectrum on the plane $z=0$ is known,
Eq.~(\ref{FieldE}) together with Eqs.~(\ref{FUV}) and (\ref{SHE})
provides the expression of the electric field vectors as
\begin{eqnarray}
E_x^{[\mathrm{W}]} &=&
\sin\theta\left(k^2+\frac{\partial^2}{\partial
x^2}\right)V_1+\cos\theta\frac{\partial^2 V_1}{\partial x \partial
z}-i\cos\theta k \frac{\partial V_2}{\partial y},\label{WEX}\\
E_y^{[\mathrm{W}]} &=& \sin\theta \frac{\partial^2 V_1}{\partial x
\partial y}+\cos\theta\frac{\partial^2 V_1}{\partial y \partial
z}-i\sin\theta k \frac{\partial V_2}{\partial z}+i\cos\theta k \frac{\partial V_2}{\partial x},\label{WEY}\\
E_z^{[\mathrm{W}]} &=&\sin\theta\frac{\partial^2 V_1}{\partial x
\partial z}+ \cos\theta\left(k^2+\frac{\partial^2}{\partial
z^2}\right)V_1+i\sin\theta k \frac{\partial V_2}{\partial
y}\label{WEZ}.
\end{eqnarray}
Here the superscript $[\mathrm{W}]$ represents the vectorial model
given by the Whittaker potentials. We find that the electric
components can be written as the Whittaker potentials and their
first- and second-order derivatives. Up to now, we have established
a general propagation model to describe the vector field structure.

It should be noted that the choice of the propagation models in the
SHEL has highly debated~\cite{Onoda2004,Bliokh2006}. It is necessary
for us to introduce the two polarization models, since optical
Magnus effect shares the same physical mechanism with SHEL. Under
the paraxial approximation of model $[\mathrm{W}]$, we can get the
two different polarization models:
\begin{eqnarray}
\tilde{\mathbf{E}}^{[\mathrm{I}]}&\propto&\bigg[\left(\alpha+\beta\frac{
k_y}{k}\cot\theta\right)\mathbf{e}_{x} +\left(\beta-\alpha \frac{k_y}{k}\cot\theta\right)\mathbf{e}_{y}\nonumber\\
&&-\frac{\alpha k_x+\beta
k_y}{k}\mathbf{e}_{z}\bigg]\tilde{f}(k_{x},k_{y})\label{modelI},
\end{eqnarray}
\begin{eqnarray}
\tilde{\mathbf{E}}^{[\mathrm{II}]}&\propto&\left(\alpha\mathbf{e}_{x}+\beta\mathbf{e}_{y}-
\frac{\alpha k_x+\beta
k_y}{k}\mathbf{e}_{z}\right)\tilde{f}(k_{x},k_{y})\label{modelII}.
\end{eqnarray}
Evidently, model $[\mathrm{II}]$ can be obtained from model
$[\mathrm{I}]$ under the condition $\theta=\pi/2$. In the context of
wave optics, both model $[\mathrm{I}]$ and model $[\mathrm{II}]$ are
exact solutions to the paraxial wave equation. However, the
customary paraxial approximation has been shown to be incompatible
with the exact Maxwell's equations~\cite{Lax1975}. The practical
light beams consist of electromagnetic fields and hence are governed
by Maxwell's equations. In recognition of this fact, it is sometimes
suggested that the model $[\mathrm{W}]$ representing the Gaussian
beam seems to be appropriate. From the experimental viewpoints, even
an ideal polarizer will transmit part of a cross-polarized
wave~\cite{Fainman1984,Aiello2009b}, and thus cannot produce a pure
linear polarization state. From the theoretical viewpoints, the
linear polarization is a paraxial approximation solution of the
Maxwell's equations. When we go beyond the paraxial approximation
and consider the lowest order corrections, the field is elliptical
polarization in the cross section~\cite{Simon1986,Simon1987}. Thus,
it is desirable to consider the mode $[\mathrm{W}]$ in the optical
Magnus effect. As shown in the following, we will compare the
results given by the three polarization models.

The changes in the transverse momentum currents can be used to
explain the physics behind the rotation phenomenon of the
wavepacket, which results in a intensity redistribution over the
wavepacket cross section~\cite{Bekshaev2005,Alexeyev2005}. The
time-averaged linear momentum density associated with the
electromagnetic field can be shown~\cite{Jackson1999} to be
\begin{equation}
\mathbf{p}^{[\mathrm{M}]}(\mathbf{r})=
\frac{1}{2c^2}\mathrm{Re}[\mathbf{E}^{[\mathrm{M}]}(\mathbf{r})
\times\mathbf{H}^{[\mathrm{M}]\ast}(\mathbf{r})]\label{LMD},
\end{equation}
Here $\mathrm{M}=\mathrm{W},\mathrm{I},\mathrm{II}$ denotes
different polarization models. The magnetic field can be obtained by
$\mathbf{H}^{[\mathrm{M}]}=-ik^{-1}
\nabla\times\mathbf{E}^{[\mathrm{M}]}$. The momentum currents can be
regarded as the combined contributions of spin and orbital parts
\begin{equation}
\mathbf{p}^{[\mathrm{M}]}=\mathbf{p}^{[\mathrm{M}]}_O+\mathbf{p}^{[\mathrm{M}]}_S.
\end{equation}
Here, the orbital term is determined by the macroscopic energy
current with respect to an arbitrary reference point and does not
depend on the polarization. The spin term, on the other hand,
relates to the phase between orthogonal field components and is
completely determined by the state of
polarization~\cite{Bekshaev2007}. In a monochromatic optical beam,
the spin and orbital currents can be respectively written in the
form
\begin{equation}
\mathbf{p}^{[\mathrm{M}]}_{S}=\mathrm{Im}[(\mathbf{E}^{[\mathrm{M}]}\cdot\nabla)\mathbf{E}^{[\mathrm{M}]\ast}],
\end{equation}
\begin{equation}
\mathbf{p}^{[\mathrm{M}]}_{O}=\mathrm{Im}[\mathbf{E}^{[\mathrm{M}]\ast}\cdot(\nabla)\mathbf{E}^{[\mathrm{M}]}],
\end{equation}
where
$\mathbf{E}^{[\mathrm{M}]\ast}\cdot(\nabla)\mathbf{E}^{[\mathrm{M}]}=E_x^{[\mathrm{M}]\ast}\nabla
E_x^{[\mathrm{M}]\ast}+E_y^{[\mathrm{M}]\ast}\nabla
E_y^{[\mathrm{M}]}+E_z^{[\mathrm{M}]\ast}\nabla E_z^{[\mathrm{M}]}$
is the invariant Berry notation~\cite{Berry2009}. It has been shown
that both spin and orbital currents originate from the beam
transverse inhomogeneity and their components are directly related
to the azimuthal and radial derivatives of the beam profile
parameters. However, the orbital currents are mainly produced by the
phase gradient, while the spin currents are orthogonal to the
intensity gradient~\cite{Bekshaev2005}. As shown in the following,
the spin and orbital currents will play different roles in the
optical Magnus effect.

\begin{figure}
\includegraphics[width=12cm]{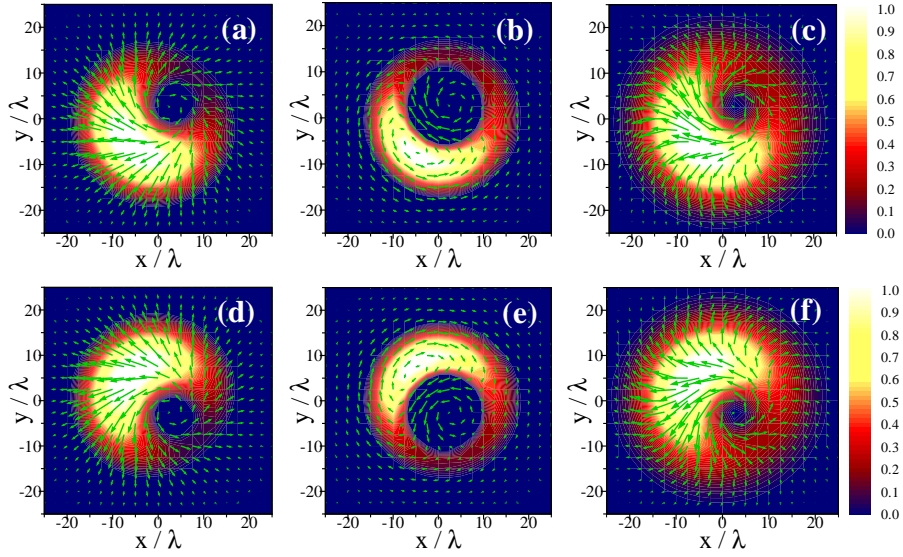}
% Here is how to import EPS art
\caption{\label{Fig2} (color online) The momentum currents for the
model $[\mathrm{W}]$ under the condition $\theta=\pi/360$. First
row: Left circularly polarized wavepacket $\sigma=+1$. Second row:
Right circularly polarized wavepacket $\sigma=-1$. The background
distribution is depicted as the longitudinal currents and the green
arrows is described as the transverse currents. [(a), (d)] Orbital
momentum currents $\mathbf{p}_{O}^{[\mathrm{W}]}$. [(b), (e)] Spin
momentum currents $\mathbf{p}_{S}^{[\mathrm{W}]}$. [(c), (f)] Total
momentum currents
$\mathbf{p}_{O}^{[\mathrm{W}]}+\mathbf{p}_{S}^{[\mathrm{W}]}$. The
cross section is chosen as $z=z_R$ and the intensity is plotted in
normalized units.}
\end{figure}

The monochromatic beam can be formulated as a localized wavepacket
whose spectrum is arbitrarily narrow~\cite{Bliokh2007}. As a
mechanical analogy, the circularly polarized wavepacket acts as the
rotating ball in our model. To generate an asymmetric intensity
distribution, we choose the angle of the fixed unit vector
$\mathbf{u}$ as $\theta=\pi/360$. Note that such a vectorial beam
should can be realized experimentally without technical
difficulties~\cite{Youngworth2000,Ren2006}. In general, the rotation
properties of wavepacket are expressed by the transverse momentum
currents as shown in Fig.~\ref{Fig2}. Very surprisingly, the orbital
currents are polarization-dependent in this polarization model
[Figs.~\ref{Fig2}(a) and \ref{Fig2}(d)]. This is due to the presence
of polarization-dependent screw wavefront. For the left circular
polarization $\sigma=+1$, the total transverse momentum currents in
the exterior part of wavepacket present an anticlockwise
circulation, while in the inner part exhibit a clockwise circulation
[Fig.~\ref{Fig2}(c)]. For the right circular polarization
$\sigma=-1$, the total transverse momentum currents present an
opposite characteristics [Fig.~\ref{Fig2}(f)]. The inherent physics
underlying this intriguing effect is the combined contributions of
transverse spin and orbital currents. The wavepacket cannot be
regarded as a rigid ball due to the opposite circulations. This is
significantly different from the mechanical Magnus effect.

\begin{figure}
\includegraphics[width=12cm]{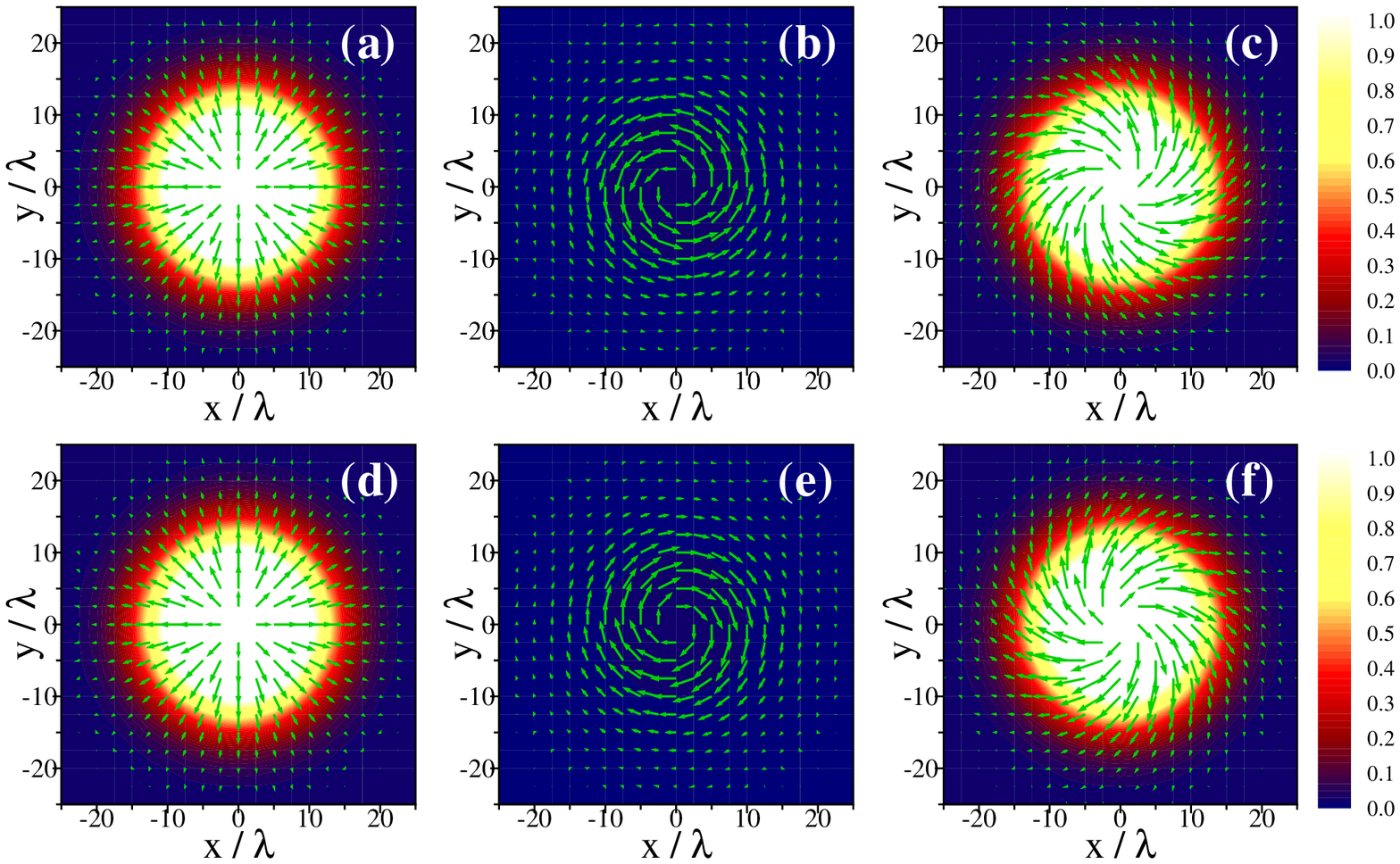}
% Here is how to import EPS art
\caption{\label{Fig3} (color online) The momentum currents for the
model $[\mathrm{W}]$ under the condition $\theta=\pi/2$. First row:
Left circularly polarized wavepacket $\sigma=+1$. Second row: Right
circularly polarized wavepacket $\sigma=-1$. The background
distribution is depicted as the longitudinal currents and the green
arrows is described as the transverse currents. [(a), (d)] Orbital
momentum currents $\mathbf{p}_{O}^{[\mathrm{W}]}$. [(b), (e)] Spin
momentum currents $\mathbf{p}_{S}^{[\mathrm{W}]}$. [(c), (f)] Total
momentum currents
$\mathbf{p}_{O}^{[\mathrm{W}]}+\mathbf{p}_{S}^{[\mathrm{W}]}$. The
cross section is chosen as $z=z_R$ and the intensity is plotted in
normalized units.}
\end{figure}

For the model $[\mathrm{W}]$ under the condition $\theta=\pi/2$,
which has also been verified by Hertz vector
method~\cite{Varga1998}, the polarization-dependent rotation of the
transverse momentum currents are plotted in Fig.~\ref{Fig3}. Very
interestingly, the orbital currents are polarization-independent in
this polarization model [Figs.~\ref{Fig3}(b) and \ref{Fig3}(e)].
However, the transverse spin currents are polarization-dependent and
the longitudinal spin currents are absent [Figs.~\ref{Fig3}(b) and
\ref{Fig3}(e)]. For the left circular polarization $\sigma=+1$, the
total transverse momentum currents present an anticlockwise
circulation [Fig.~\ref{Fig3}(c)]. For the right circular
polarization $\sigma=-1$, the total transverse momentum currents
present a clockwise circulation [Fig.~\ref{Fig3}(f)]. The transverse
intensity distribution of this model is axially symmetric and the
polarization-dependent rotation, if it exists, cannot be observed
directly. This is a possible reason why the polarization-dependent
split is observed only on the plane which is not perpendicular to
the propagation direction of the beam~\cite{Aiello2009a}.

\section{Optical Magnus Effect}\label{SecIII}
To illustrate the polarization-dependent rotation effect, we now
determine the shift of wavepacket centroid, which is given by
$\langle\mathbf{r}\rangle^{[\mathrm{M}]}=\langle
x\rangle^{[\mathrm{M}]} \mathbf{e}_{x} + \langle y
\rangle^{[\mathrm{M}]} \mathbf{e}_{y}$, with
\begin{equation}
\langle \mathbf{r}\rangle^{[\mathrm{M}]} = \frac{\int \int
\mathbf{r} p_z^{[\mathrm{M}]}(x,y,z) \text{d}x \text{d}y}{\int \int
p_z^{[\mathrm{M}]}(x,y,z) \text{d}x \text{d}y}\label{Centroid}.
\end{equation}
First let us examine the polarization-dependent rotation in the
model $[\mathrm{W}]$. By substituting Eqs.~(\ref{WEX}), (\ref{WEY})
and (\ref{LMD}) into Eq.~(\ref{Centroid}), we obtain the shifts as
\begin{equation}
\langle x \rangle^{[\mathrm{W}]}=-\frac{z
\sin2\theta}{2(\cos^2\theta+k z_R\sin^2\theta)}\label{CentroidXW},
\end{equation}
\begin{equation}
\langle y \rangle^{[\mathrm{W}]}=-\frac{\sigma z_R
\sin2\theta}{2(\cos^2\theta+k z_R\sin^2\theta)}\label{CentroidYW},
\end{equation}
where $z_R=k w_0^2/2$ is the Rayleigh length. We find that the
wavepacket centroid shifts a distance away from the propagation axis
in $xoy$ plane. The shift $\langle x \rangle^{[\mathrm{W}]}$ is
polarization independent, while the shift $\langle y
\rangle^{[\mathrm{W}]}$ depends on the polarization state $\sigma$.
Note that the shift $\langle x \rangle^{[\mathrm{W}]}$ can be
regarded as a small angle inclining from the propagation
axis~\cite{Merano2009,Luo2009}. However, the shift $\langle y
\rangle^{[\mathrm{W}]} $ does not change on propagation.

The angular positions of the wavepacket centroid indicates the
rotation angle $\tan\varphi^{[\mathrm{W}]}=\langle x
\rangle^{[\mathrm{W}]}/\langle y \rangle^{[\mathrm{W}]}$, which is
significantly different from the previous
definition~\cite{Bekshaev2006}. According to Eq.~(\ref{Centroid})
the rotation angle is given by
\begin{equation}
\varphi^{[\mathrm{W}]}= \arctan \frac{z}{\sigma z_R}.\label{Angle}
\end{equation}
Very surprisingly, the rotation angle is proportional to the Gouy
phase for circular polarization $\sigma=\pm 1$. In all cross
sections denoted by the distance from the $z=0$ plane, the
transverse patterns are quite similar. However the overall scale
changes and the pattern rotates as a whole. The former is caused by
diffraction, and the latter is seen due to the existence of
transverse momentum currents.

Since the rotating wavepacket passes the distance between different
cross-sections in time $t=z/c$, the instant angular velocity can be
given by  $\Omega^{[\mathrm{W}]}(z)=d \varphi^{[\mathrm{W}]} / d t=c
d\varphi^{[\mathrm{W}]} / d z $, so that
\begin{equation}
\Omega^{[\mathrm{W}]}(z)=c\frac{\sigma z_R}{\sigma^2
z_R^2+z^2}\label{AngleV}.
\end{equation}
The rotation velocity of the wavepacket centroid decreases as the
propagation distance increases. When the condition $\sigma=0$ is
satisfied and the rotation characteristics of wavepacket centroid
vanish, i.e., $\Omega^{[\mathrm{W}]}(z)=0$. This is why the linear
polarized wavepacket cannot present the optical Magnus effect.
However the linear polarization can be represented as a
superposition of two circularly polarized
components~\cite{Bliokh2006}. As the result, the
polarization-dependent split of the wavepacket intensity
distribution arises. Thus, the same mechanism also leads to other
effects such as the SHEL.

\begin{figure}
\includegraphics[width=12cm]{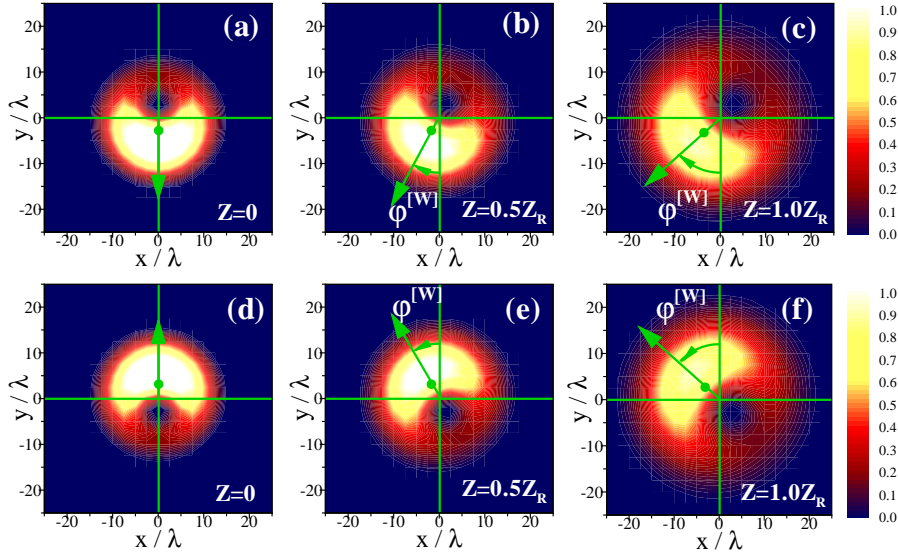}
% Here is how to import EPS art
\caption{\label{Fig4} (color online) The polarization-dependent
rotations of wavepacket centroid (green point) for the model
$[\mathrm{W}]$. First row: The wavepacket centroid undergoes a
clockwise rotation for left circular polarization $\sigma=+1$.
Second row: The wavepacket centroid presents an anticlockwise
rotation for right circular polarization $\sigma=-1$. Images are
labeled by corresponding values of propagation length $z$ and
rotation angle $\varphi^{[\mathrm{W}]}$. The intensity distributions
are plotted in the plane [(a), (d)] $z=0$, [(b), (e)] $z=0.5z_R$,
and [(c), (f)] $z=z_R$. Other parameters are the same as those in
Fig.~\ref{Fig2}.}
\end{figure}

Figure~\ref{Fig4} shows the polarization-dependent rotations of
wavepacket centroid. At the plane $z=const.$, the intensity of the
wavepacket can be regarded as
$I^{[\mathrm{W}]}(\mathbf{r})=c^2p^{[\mathrm{W}]}_z(\mathbf{r})$.
The wavepacket centroid (green spots) whose angular positions
indicate the rotation angle $\varphi$. For the left circular
polarization $\sigma=+1$, the wavepacket centroid exhibits a
clockwise rotation [Figs.~\ref{Fig4}(a)-\ref{Fig4}(c)]. For the
right circular polarization $\sigma=-1$, however, the wavepacket
centroid presents an anticlockwise rotation
[Figs.~\ref{Fig4}(d)-\ref{Fig4}(f)]. During the wavepacket
propagation from $z=0$ to $z=+z_R$, the rotation angle amounts to
$\varphi^{[\mathrm{W}]}=\sigma\pi/4$. The physics underlying this
intriguing effect is the combined contributions of transverse spin
and orbital currents. The novel polarization-dependent rotations
differs from the conventional optical Magnus effect, in that
light-matter interaction is not required.

We are currently investigating the polarization mode $[\mathrm{I}]$.
On substituting Eqs.~(\ref{ASR}) and (\ref{modelI}) into
Eq.~(\ref{Centroid}) we have
\begin{equation}
\langle x \rangle^{[\mathrm{I}]}=0~~~~~\langle y
\rangle^{[\mathrm{I}]} =-\frac{\sigma \cot\theta}{k
}\label{CentroidI}.
\end{equation}
There exists an inherent transverse shift on propagation. This
result coincides with that obtained by Li~\cite{Li2009a} with
different methods. According to Eq.~(\ref{CentroidI}) the rotation
angle is given by
\begin{equation}
\varphi^{[\mathrm{I}]}= \sigma\arctan \frac{\pi}{2}.\label{Angle}
\end{equation}
In this case, the rotation characteristics of wavepacket centroid
vanish, i.e., $\Omega^{[\mathrm{I}]}(z)=0$.

\begin{figure}
\includegraphics[width=12cm]{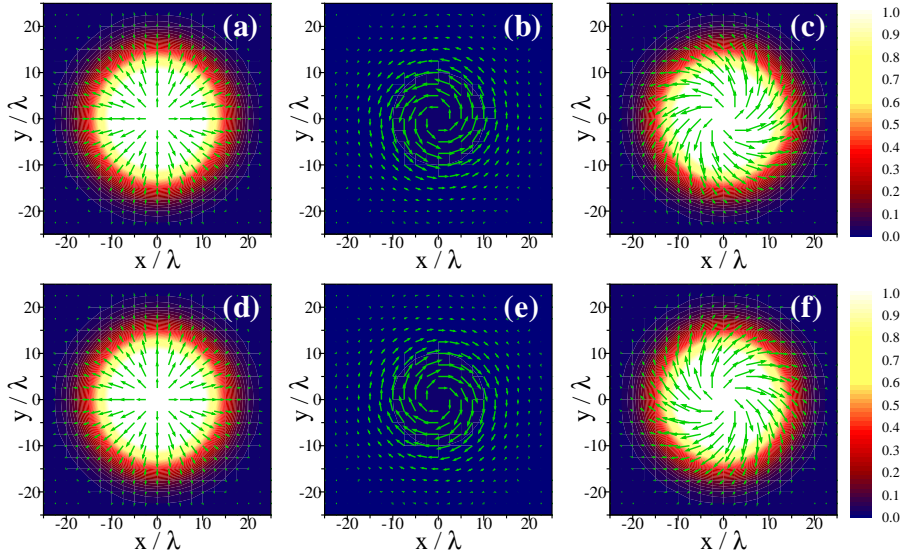}
% Here is how to import EPS art
\caption{\label{Fig5} (color online) The polarization-dependent
rotations of transverse momentum currents for the model
$[\mathrm{II}]$. First row: Left circularly polarized wavepacket
$\sigma=+1$. Second row: Right circularly polarized wavepacket
$\sigma=-1$. The background distribution is depicted as the
longitudinal currents and the green arrows is described as the
transverse currents. [(a), (d)] Orbital momentum currents
$\mathbf{p}_{O}^{[\mathrm{II}]}$. [(b), (e)] Spin momentum currents
$\mathbf{p}_{S}^{[\mathrm{II}]}$. [(c), (f)] Total momentum currents
$\mathbf{p}_{O}^{[\mathrm{II}]}+\mathbf{p}_{S}^{[\mathrm{II}]}$. The
cross section is chosen as $z=z_R$ and the intensity is plotted in
normalized units.}
\end{figure}

We are now in a position to consider the polarization model
$[\mathrm{II}]$. On substituting Eqs.~(\ref{ASR}) and
(\ref{modelII}) into Eq.~(\ref{Centroid}) we can determine
\begin{equation}
\langle x \rangle^{[\mathrm{II}]}=0,~~~~~\langle y
\rangle^{[\mathrm{II}]}=0\label{CentroidII}.
\end{equation}
A further important point should be noted is that
Eq.~(\ref{CentroidII}) can be obtained from Eq.~(\ref{CentroidI})
under the condition $\theta=\pi/2$. In this model, the
polarization-dependent rotation of the centroid cannot be observed
directly. Thus, we attempt to explore an alternative way to describe
the polarization-dependent rotation effect. When the momentum
currents are included, the change rate of azimuthal angle with $z$
axis is written as~\cite{Padgett1995}
\begin{equation}
\frac{\partial\varphi^{[\mathrm{II}]}}{\partial
z}=\frac{p_\varphi^{[\mathrm{II}]}}{r
p_z^{[\mathrm{II}]}}.\label{AngleM}
\end{equation}
Here, $p_\varphi$ describes the momentum current that circulates
around the propagation axis, and $p_z$ describes the momentum
current that propagates along the $+z$ axis. The result of
Eq.~(\ref{LMD}) can be expressed in term of azimuthal component,
defined by
\begin{equation}
p_\varphi^{[\mathrm{II}]}=-p_x^{[\mathrm{II}]}\sin\varphi
+p_y^{[\mathrm{II}]}\cos\varphi\label{PVAR},
\end{equation}
By substituting Eqs.~(\ref{LMD}) and (\ref{PVAR}) into
Eq.~(\ref{AngleM}) and carrying out the integration, we obtain
\begin{equation}
\varphi^{[\mathrm{II}]}= \sigma \arctan \frac{z}{z_R}\label{Angle}.
\end{equation}
This result is slightly different from the rotation of the centroid
in model $[\mathrm{W}]$. The instant angular velocity can be given
by $\Omega^{[\mathrm{II}]}(z)=c d\varphi^{[\mathrm{II}]}/ d z $, so
that
\begin{equation}
\Omega^{[\mathrm{II}]}(z)=c \frac{\sigma z_R}{
z_R^2+z^2}\label{AngleV}.
\end{equation}
The rotation velocity of the momentum current decreases as the
propagation distance increases. When the condition $\sigma=0$ is
satisfied, the rotation characteristics vanish.

To obtain a clear physical picture, the polarization-dependent
rotations of the momentum currents are plotted in Fig.~\ref{Fig5}.
We find that the orbital currents are polarization-independent in
this polarization model [Figs.~\ref{Fig5}(a) and \ref{Fig5}(d)].
However, the spin currents are polarization-dependent
[Figs.~\ref{Fig5}(b) and \ref{Fig5}(e)]. For the left circular
polarization $\sigma=+1$, the total transverse momentum currents
present an anticlockwise circulation [Fig.~\ref{Fig5}(c)]. For the
right circular polarization $\sigma=-1$, the total transverse
momentum currents present a clockwise circulation
[Fig.~\ref{Fig5}(f)]. The polarization-dependent rotations of
momentum currents may provide an alternative way to illustrate the
optical Magnus effect, although whether the momentum currents in the
free space propagating along a curvilinear trajectory is still
debated~\cite{Allen2000,Volyar1999}. Comparing Fig.~\ref{Fig3} with
Fig.~\ref{Fig5} shows that there are no notable difference between
the model $[\mathrm{W}]$ (under the condition $\theta=\pi/2$) and
the model $[\mathrm{II}]$. In fact, the latter can be regarded as
the paraxial approximation of the former.

\section{Spin and orbital angular momenta}\label{SecII}
In the frame of classical electrodynamics, it is the circularly
polarized wavepacket but not the spin photon acting as the rotating
ball. Now a question arises: What roles the spin and orbital angular
momenta play in the optical Magnus effect? We now in the position to
analysis the angular momentum density for each of individual
wavepacket, which can be written as~\cite{Jackson1999}
\begin{equation}
\mathbf{j}^{[\mathrm{M}]}(\mathbf{r})=\mathbf{r}\times\mathbf{p}^{[\mathrm{M}]}(\mathbf{r})\label{AMD}.
\end{equation}
Within the paraxial approximation, the angular momentum can be
divided into the spin and orbital angular parts
$\mathbf{j}^{[\mathrm{M}]}=\mathbf{j}^{[\mathrm{M}]}_O+\mathbf{j}^{[\mathrm{M}]}_S$~\cite{Allen1992},
it follows that
\begin{equation}
\mathbf{j}^{[\mathrm{M}]}_O=\mathbf{r}\times\mathbf{p}^{[\mathrm{M}]}_O,
\end{equation}
\begin{equation}
\mathbf{j}^{[\mathrm{M}]}_S=\mathbf{r}\times\mathbf{p}^{[\mathrm{M}]}_S.
\end{equation}
This separation should hold beyond the paraxial
approximation~\cite{Barnett2002}.

We first consider the longitudinal angular momentum density $j_z$
which can be regarded as the combined contributions of spin and
orbital parts:
\begin{equation}
j^{[\mathrm{M}]}_\mathrm{Oz}=x p^{[\mathrm{M}]}_{Oy}-y
p^{[\mathrm{M}]}_{Ox},
\end{equation}
\begin{equation}
j^{[\mathrm{M}]}_{Sz}=x p^{[\mathrm{M}]}_{Sy}-yp^{[\mathrm{M}]}
_{Sx}.
\end{equation}
The longitudinal angular momentum will provide a simple way to
understand why the circularly polarized wavepacket exhibits the
optical Magnus effect.

\begin{figure}
\includegraphics[width=12cm]{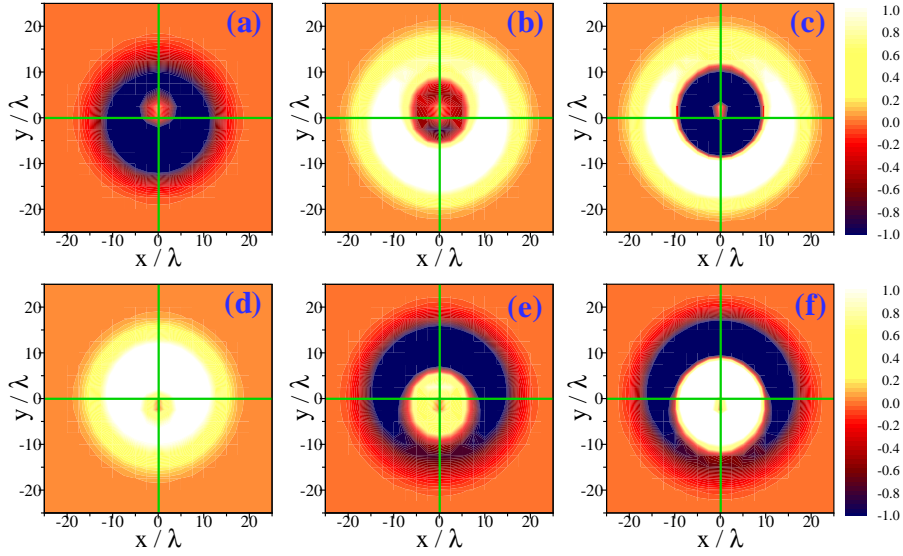}
% Here is how to import EPS art
\caption{\label{Fig6} (color online) The distribution of the
longitudinal angular momentum density for the model $[\mathrm{W}]$.
First row: Left circular polarization $\sigma=+1$. Second row: Right
circular polarization $\sigma=-1$. [(a), (d)] Orbital angular
momentum density $j_{Oz}^{[\mathrm{W}]}$. [(b), (e)] Spin angular
momentum density $j_{Sz}^{[\mathrm{W}]}$. [(c), (f)] Total angular
momentum density $j_{Oz}^{[\mathrm{W}]}+j_{Sz}^{[\mathrm{W}]}$. The
cross section is chosen as $z=0$ and the intensity is plotted in
normalized units. Other parameters are the same as those in
Fig.~\ref{Fig2}.}
\end{figure}

Figure~\ref{Fig6} presents the distribution of longitudinal angular
momentum density for the model $[\mathrm{W}]$.  Very surprisingly,
this polarization model of wavepacket possesses
polarization-dependent orbital angular momentum density
[Figs.~\ref{Fig6}(a) and \ref{Fig6}(d)]. However, the spin momentum
density exhibits a significantly different distribution
[Figs.~\ref{Fig6}(b) and \ref{Fig6}(e)]. For the left circular
polarization $\sigma=+1$, the total angular momentum density in the
exterior part of wavepacket is positive $j_z>0$, while in the inner
part is negative $j_z<0$ [Fig.~\ref{Fig6}(c)]. This means that the
outer part of the packet presents an anticlockwise rotation, the
inner part undergoes a clockwise one. For the right circular
polarization $\sigma=-1$, the total angular momentum density of
wavepacket presents an oppositive distribution [Fig.~\ref{Fig6}(f)].
Thus, the rotating wavepacket cannot be regarded as a rigid ball as
in the mechanical Magnus effect. This is the reason why we choose
the centroid as the reference point to describe the
polarization-dependent trajectories.

\begin{figure}
\includegraphics[width=12cm]{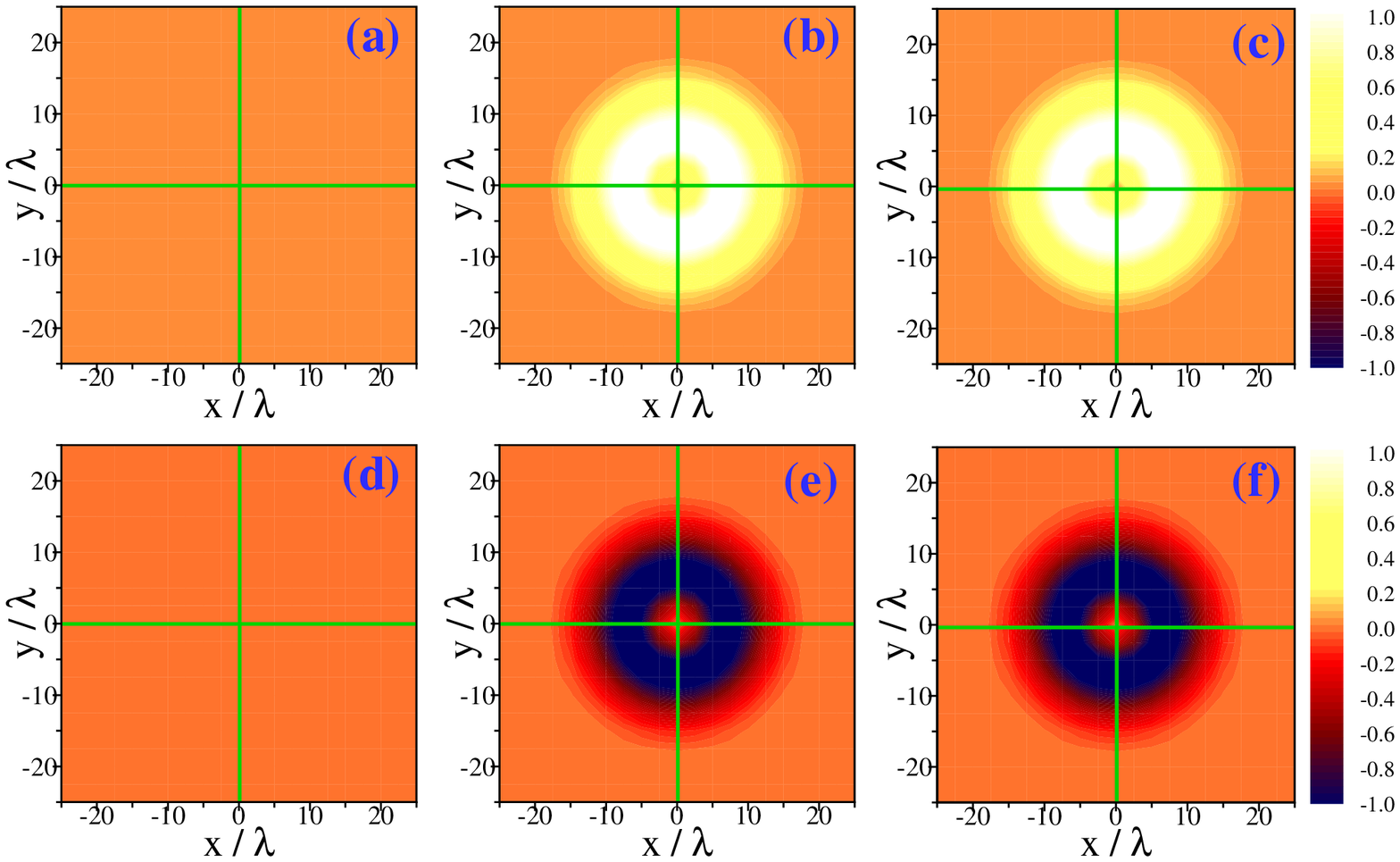}
% Here is how to import EPS art
\caption{\label{Fig7} (color online) The distribution of the
longitudinal angular momentum density for the model $[\mathrm{II}]$.
First row: Left circular polarization $\sigma=+1$. Second row: Right
circular polarization $\sigma=-1$. [(a) and (d)] Orbital angular
momentum density $j_{Oz}^{[\mathrm{II}]}$. [(b) and (e)] Spin
angular momentum density $j_{Sz}^{[\mathrm{II}]}$. [(c) and (f)]
Total angular momentum density
$j_{Oz}^{[\mathrm{II}]}+j_{Sz}^{[\mathrm{II}]}$. The cross section
is chosen as $z=0$ and the intensity is plotted in normalized units.
Other parameters are the same as those in Fig.~\ref{Fig5}.}
\end{figure}

For the comparison, we plot the longitudinal momentum density of the
model $[\mathrm{II}]$ in Fig.~\ref{Fig7}. By comparing with the
model $[\mathrm{W}]$, we find that the orbital angular momentum
density vanishes in the present polarization model
[Figs.~\ref{Fig7}(a) and \ref{Fig7}(d)]. Thus, only the spin
momentum density exhibits a polarization-dependent distribution
[Figs.~\ref{Fig7}(b) and \ref{Fig7}(e)]. For the left circular
polarization $\sigma=+1$, the total angular momentum density first
increases then decreases with the increase of $r$ as shown in
Fig.~\ref{Fig7}(c). For the right circular polarization $\sigma=-1$,
the total angular momentum density presents an oppositive
distribution as shown in Fig.~\ref{Fig7}(f). The transverse
intensity distribution is axially symmetric and the
polarization-dependent rotation, if it exists, cannot be observed
directly.

Now we want to enquire what roles the angular momentum play in the
optical Magnus effect. To answer this question needs to discuss the
transverse angular momentum. The time-averaged linear momentum and
angular momentum, which can be obtained by integrating over the
whole $x-y$ plane~\cite{Jackson1999}
\begin{equation}
\mathbf{P}^{[\mathrm{M}]}=\int \int
\mathbf{p}^{[\mathrm{M}]}(\mathbf{r})\text{d}x \text{d}y,
\end{equation}
\begin{equation}
\mathbf{J}^{[\mathrm{M}]}=\int\int
\mathbf{j}^{[\mathrm{M}]}(\mathbf{r})\text{d}x \text{d}y\label{LAM}.
\end{equation}
The transverse angular momentum components are given by
\begin{equation}
J^{[\mathrm{M}]}_{x}=\langle y \rangle P^{[\mathrm{M}]}_{z}-z
P^{[\mathrm{M}]}_{y},
\end{equation}
\begin{equation}
J^{[\mathrm{M}]}_{y}=z P^{[\mathrm{M}]}_{x}-\langle x \rangle
P^{[\mathrm{M}]}_{z}\label{TAM}.
\end{equation}
In the $z=0$ plane, we have
$J_{x}^{[\mathrm{M}]}/P_{z}^{[\mathrm{M}]}=\langle y
\rangle^{[\mathrm{M}]}$ and
$J_{y}^{[\mathrm{M}]}/P_{z}^{[\mathrm{M}]}=-\langle x
\rangle^{[\mathrm{M}]}$. Thus, the transverse angular momentum can
be obtained by measuring the position of wavepacket
centroid~\cite{Aiello2009a}. In order to reveal the rotation
characteristics, it is necessary for us to know the transverse
angular momentum in any cross section. We first consider the model
$[\mathrm{W}]$, the transverse angular momenta are given by
\begin{equation}
J^{[\mathrm{W}]}_{x}=-\frac{\pi \sigma \sin2\theta}{2 k^2 z_R},~~~~~
J^{[\mathrm{W}]}_{y}=\frac{\pi z \sin2\theta}{2 k^2
z_R^2}\label{CentroidY}.
\end{equation}
In this case, the polarization-dependent rotation of the centroid is
unavoidable since the wavepacket possesses transverse angular
momentum. For the model $[\mathrm{I}]$, we obtain
\begin{equation}
J^{[\mathrm{I}]}_{x}=-\frac{\pi \sigma \cot\theta}{k^2 z_R},~~~~~~
J^{[\mathrm{I}]}_{y}=0\label{TAM}.
\end{equation}
This is the reason why the model $[\mathrm{I}]$ only exhibits a
transverse shift. For the model $[\mathrm{II}]$, however
\begin{equation}
J^{[\mathrm{II}]}_{x}=0,~~~~~~ J^{[\mathrm{II}]}_{y}=0\label{TAM}.
\end{equation}
Thus, the wavepacket centroid no longer presents the
polarization-dependent rotation. The physics underlying this
phenomenon is the absence of the transverse angular momentum. It
should be noted that the SHEL can be noticeably enhanced when the
wavepacket carries orbital angular
momentum~\cite{Bliokh2009a,Bliokh2009b,Fadeyeva2009}. Further work
is needed to uncover the optical Magnus effect of such a wavepacket
in the free space.

It should be mentioned that the recent advent of negative index
metamaterials, also known as left-handed materials
(LHMs)~\cite{Veselago1968,Shelby2001}, can induce a reversed
polarization-dependent rotation of the trajectory of the wavepacket
centroid. Because of the negative index, we can expect a negative
Rayleigh length in LHMs~\cite{Luo2008a,Luo2008b}. It will be
interesting for us to describe in detail how the wavepacket
trajectory evolves in the LHMs. Recently, the technique of
transformation optics has emerged as a means of designing
metamaterials that can bring about unprecedented control of
electromagnetic fields~\cite{Pendry2006}. It is possible that the
trajectories of circularly polarized wavepacket can be controlled by
introducing a prescribed spatial variation in the constitutive
parameters.

\section{Conclusions}
In conclusion, we have established a general vector field model to
describe the role of transverse momentum currents in optical Magnus
effect in the free space. We have demonstrated the existence of a
novel optical polarization-dependent Magnus effect which differs
from conventional optical Magnus effect in that light-matter
interaction is not required. In the optical Magnus effect, the
circularly polarized wavepacket acts as the rotating ball, but is
not identical to a rigid body. This is because different parts of
the wavepacket present diverse rotation characteristics. For a
certain circularly polarized wavepacket, whether the rotation is
clockwise or anticlockwise depends on the polarization state. Such a
polarization-dependent rotation is unavoidable when the wavepacket
possesses transverse momentum currents. We predict that this novel
effect may be observed experimentally even in the propagation
direction. Our findings provide further evidence for the optical
Magnus effect in the free space. Because of the close similarity in
atom physics, condensed matter, and optical physics, we believe that
the Magnus effect is not limited to electromagnetic fields, but
extends to other research areas, such as atom, ion, and electron
beams.

\begin{acknowledgements}
We are sincerely grateful to the anonymous referee, whose comments
have led to a significant improvement of our paper. This research
was supported by the National Natural Science Foundation of China
(Grants Nos. 10804029, 10904036, 60890202, and 10974049).
\end{acknowledgements}

\end{document}